# Survey on Some Optimization Possibilities for Data Plane Applications

Gereltsetseg Altangerel, Tejfel Máté

Department of Programming Languages and Compilers, ELTE,
Eötvös Loránd University, Budapest, Hungary

## Abstract

*By programming both the data plane and the control plane, network operators can customize their networks based on their needs, regardless of the hardware manufacturer. Control plane programming, a major component of the SDN (Software Defined Network) concept, has been developed for more than 10 years and successfully implemented in real networks. Efforts to develop reconfigurable data planes and high-level network programming languages make it truly possible to program data planes. Therefore, the programmable data planes and SDNs offer great flexibility in network customization, allowing many innovations to be introduced on the network. The general focus of research on the data plane is data-plane abstractions, languages and compilers, data plane algorithms, and applications. This paper outlines some emerging applications on the data plane and offers opportunities for further improvement and optimization.*

## 1. Introduction

Efforts to create fully programmable networks have been going on for a very long time. SDN technology is part of this long history. Many factors in business and technology have led to the creation of SDNs and fully programmable networks. In the case of a technological view, the main difficulty of traditional networks is a proprietary system, so the introduction of new technologies and protocols is very slow, and network management is complicated. From a business perspective, network operators needed to reduce investment and operation costs, and take full control of their networks: defining their own control plane and data plane algorithms.

To meet these requirements, the SDN defines two important features: first, separate the data plane and control panel, and second, the controller platforms can control multiple forwarding elements using a well-defined API (Application Programming Interface), such as OpenFlow, one of the successful protocols. Not only do they simplify network management, but they also open up development gateways to each control plane and data plane, allowing for many network innovations.

So far, SDN has made significant strides in the industry and has conducted extensive research on the control plane. For instance, commercial switches have supported OpenFlow, and a variety of controller platforms have emerged, based on which several control plane applications have developed and implemented in the major data centers such as Google.

A fully programmable network has two pillars: a programmable data plane and a programmable control plane. The data plane programming began to be discussed in the 2000s with the advent of the merchant chip, but it became a reality in 2015, and research in this area is in great demand today [1].

In recent years, researches on developing data plane programming languages, creating programmable switching architecture, and improving the performance of programmable switches have become more mature. Thanks to the success of these fundamental studies and implementations, data plane applications are evolving, and some are entering production and beginning to bear fruit.

The general focus of research on the data plane is data-plane abstractions, languages and compilers, data plane algorithms, and applications [2]. This paper summarizes the research of some data plane applications and provides some ideas in detail on how to improve and optimize these applications.

The rest of this survey paper is organized as follows. Section 2 provides background information on data plane programming, Section 3 describes the data plane applications and future optimization ideas, and the final section presents conclusions and future work.

## 2. BACKGROUND

### 2.1. Data plane programmability

Network devices process packets with the help of control plane and data plane algorithms. Data plane algorithms define the forwarding behaviour of a network device (packet processing stages, tables, and so on), while control plane algorithms define rules for manipulating a packet in the data plane, sense network, detect network failures, and update packet processing rules [3]. In the SDN network, the control plane algorithms running on the controller platform (e.g., server) manage the data plane. For example, routing algorithms in the control plane define packet forwarding rules based on the destination IP address. These rules are installed in the routing table of the data plane via API.

As a result of many years of research and improvement on SDN, the control plane can be flexibly programmed by the end-user (network operator). The data plane also needed to be programmed by the end-user to introduce innovations quickly in the network. In a traditional network, the implementation process of a new feature [4] or protocol goes through many stages, from software developer to standard organization and chip designer. For example, as a result of this process, it took 4.5 years for the VxLAN protocol to the network from the first proposal [5]. This can be seen as one of the critical reasons why internet architecture has not changed for many years. If the data plane is flexible enough, both the end-user and hardware vendor can quickly deploy the new features [3].

The following research works have been taken to make the data plane programmable:

1. **Developing data plane programming languages**: Domain-specific programming languages for defining data plane algorithms and functionalities (forwarding behaviour) are being developed. Examples include FAST [6], Domino [7], Protocol-Oblivious Forwarding [8], and NetKAT [9], P4 [10], with P4 being the most successful.

2. **Creating data plane architectures**: To map the data plane algorithm defined by domain-specific language to the switch ASIC hardware, the hardware vendor must provide the data plane architecture (programmable building blocks and data plane interface between them)[11]. This architecture is also called a data plane model or hardware abstraction in some literature. For example, architecture for Tofino programmable switches is protocol independent switching architecture (PISA) and based on this architecture, data plane algorithms can be defined in P4.

3. **Improving performance of programmable switch**: The belief that the performance of a programmable switch cannot reach the performance of a fixed-function switch is obsolete, and thanks to numerous studies and technologies, the performance of a programmable switch has approached/same as that of a fixed-function switch [12].

4. **Defining API**: Providing an interface for connecting the control plane and the programmable data plane. For example, the P4 compiler creates an API that connects the data plane to the control plane [11].

## 2.2. P4 language

P4 is a domain-specific programming language for defining packet processing algorithms on the data plane of a programmable network device (target)[11]. This section briefly describes the P4 language, its advantages, programmable packet processing components that can be defined in P4, and how to run P4 code on the target.

P4 is currently one of the most popular and well-defined languages and has two main advantages: target and protocol-independency, so a lot of data plane applications currently in development are on P4. Target-independency means that P4 program can run on any type of target. To ensure this feature, the hardware vendor must implement a generic architecture and a compiler backend for a given target: provide them to the P4 developer [13] and so, the P4 program is easily mapped to the target with help of these. Protocol-independency means that P4 developers can define their rich set of protocols and data plane behaviour/functionalities.

According to the architecture, the main blocks that can be programmed on P4 are packet parser, match-action units (one or more), and deparser. The general data plane architectures of P4 for research purposes are V1 and Portable Switch Architecture (PSA) [14] and, optimization of the P4 data plane application has been doing based on these architectures. Figure 1 describes basic pipeline in V1 model architecture. Parser recognizes incoming packets and extracts headers and fields from the packet. After this, the match-action pipeline processes extracted packet headers. A match-action unit contains one or more tables and actions. For example, the IPv4 routing table showed in Figure 2 can be created here, and the match key is the destination IP address and based on which, corresponding actions such as drop or forward are performed. In this stage, the header can be added, subtracted, and modified. The deparser builds the outgoing packet by assembling the processed headers and the original packet payload [10]. In the case of PSA architecture, it is possible to define more detailed pipelines with more than one pair of parser and deparser for ingress and egress. Also, the match-action tables and external functions can be determined between parser and deparser.

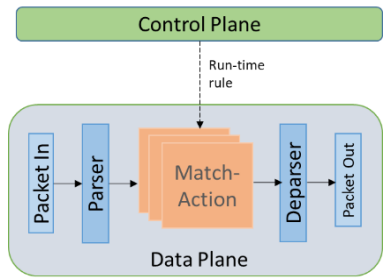

Figure 1. Abstract packet forwarding in P4

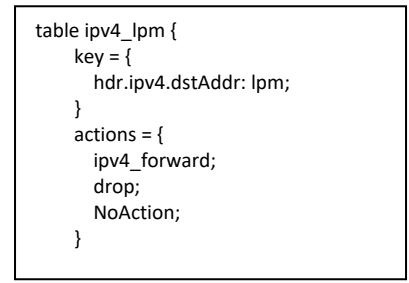

Figure 2. IPv4 table example

## 3. DATA PLANE APPLICATIONS AND OPTIMIZATIONS

Developing novel and optimal applications on the data plane are one of the interesting areas of data plane research. The general directions of applications are in-band telemetry, load-balancing, in-network computation, deployment of consensus protocols, congestion control, queue management and traffic management [2]. This section describes some of these applications, their motivation, approach, challenges, and future improvement and optimization possibilities.

### 3.1. In-band Network Telemetry (INT)

One of the foundations for effective network management and control is network measurement. More accurate, precise, real-time measurements are considered good. Traditional measurement and monitoring methods are active methods (ping, traceroute),  passive methods based on traffic mirrors, and hybrid method -a combination of these [15]. These are simple to deploy, but the downside is that they put extra load on the network during monitoring, so they can't be much precise in some cases. In other words, the process of measuring the network itself can affect the state of the network.

With the advent of programmable data planes, the In-band Network Telemetry framework, a more direct network measurement, is originated on a data plane without the involvement of a control plane. The basic idea is to collect the status of network devices (metadata) using a normal packet or probe packet (INT packet) that is transmitted over the network. Intermediary devices embed their own metadata into the INT packet. Therefore, it does not create a much more additional load on the network compared with traditional measurement. Also, it is more detailed, accurate and near-real time. One disadvantage is that metadata is limited by packet's maximum transmission unit (MTU). INT instructions (header) on what to collect from the devices are added to packet at the source INT node and then that packet is transmitted through network for collecting device's state. The metadata and INT header is removed from packet on the edge device (INT sink node). The sink node then performs the appropriate monitoring or actions, for example, it forwards the collected report to another external device or server for further monitoring [16].

INT operation can be divided into 3 phases:

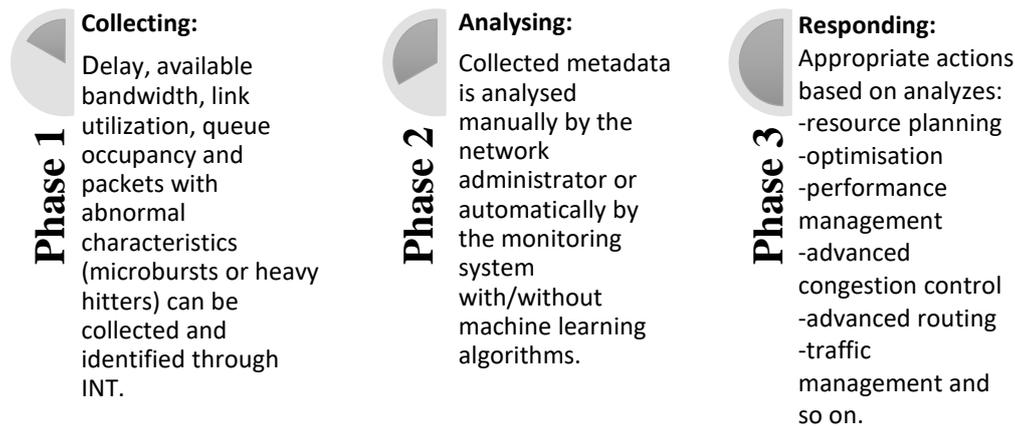

**Phase 1**

**Collecting:**

Delay, available bandwidth, link utilization, queue occupancy and packets with abnormal characteristics (microbursts or heavy hitters) can be collected and identified through INT.

**Phase 2**

**Analysing:**

Collected metadata is analysed manually by the network administrator or automatically by the monitoring system with/without machine learning algorithms.

**Phase 3**

**Responding:**

Appropriate actions based on analyzes:
-resource planning
-optimisation
-performance management
-advanced congestion control
-advanced routing
-traffic management and so on.

Figure 3. INT phases

### 3.1.1.  Recent optimization works around INT

**Optimization on phase 1:** The most important thing to consider when collecting network status via INT is not to compromise the performance of the network, intermediary devices, and the monitoring servers. To fulfill this requirement, it is important to determine the proper size and structure of the INT packet within the MTU, minimize the number of flows/packets for telemetry, filter unimportant telemetry information, and choose optimal collection mechanisms. Some studies around these are:

- **Optimizing telemetry data:** It can be optimizing number of monitored network elements, not including the INT header in all flow packets [17], sampling packets for monitoring, and using threshold to identify in-band monitored flow. For example, P4-based INT doesn't support sampling; therefore, adding an INT header to all incoming packets will create high network overhead in a large scale network. One of the works on this [18] suggests sampling strategies based on rate and event. In the rate based strategy, the INT source node inserts the INT header into every Rth packet, where R is a configurable parameter.  Another work [18] bound the amount of information added to each packet.
- **Intelligent trigger:** Fault detection platform with event-based and policy-based trigger is considered intelligent mechanism. The event can be detected by data plane or monitoring server. This solution [17] offloads event detection from monitoring server to an in-network P4 application and it reduces network overhead and monitoring server load. KeySight [19] suggested event-triggered fault detection platform based on Bloom filter. PAINT [20] offers policy based detection (by monitoring system): network operators use Service Provisioning Language (SPL) to define and deploy in-band network telemetry services. PAINT automatically parses service policies.

Since the network device and the monitoring node (server, sink, and analyser) both have limited processing capabilities, it is important to determine whether it is optimal to detect the event with either a network device or a monitoring server. In addition, event-based and policy-based detection algorithms themselves are one thing to optimize.

**Optimization on phase 2:** To make monitoring more effective, it is being combined with machine learning methods. The machine learning part of the monitoring system is called the knowledge plane in some works [21] or the knowledge-defined network. An example of these solutions is Barefoot Deep Insight [22], which is the world's first commercial-level packet-by-packet status monitoring system. Combined with machine learning technologies, Barefoot Deep Insight can realize automatic abnormality monitoring of network performance at nanosecond time, including microburst detection, abnormal packet loss detection, abnormal queue detection, and so on.

**Optimization on phase 3:** The applications such as congestion control and advanced routing based on INT should be efficient. For example, the efficiency of congestion control is evaluated based on the congestion detection, and the resolving time. There are congestion control mechanisms on different INT characteristics such as link load based [23],   rate-based [24], and queuing and processing delay based congestion control [25]. The efficiency of these solutions should be determined by performance evaluation.

Therefore, in addition to the optimization ideas mentioned at each of the above INT stages, there are opportunities for further research to develop solutions for other types of networks, such as adapting or expanding the in-band telemetry system in a wireless sensor network (WSN) and Internet of Things (IoT) data network. For example, IoT packets are too small, making it difficult to identify abnormal behaviour of packet [26].

## 3.2. In-Network Computing

Traditional network devices often focus on achieving high throughput, so processing is limited on transmitted data. With the advent of flexible programmable networking devices, it has become possible to perform more computations on network devices (in-network computations) without reducing packet processing rates. In other words, it means that a set of computing operations on an end-server and middlebox can be offloaded to a network device. This has the following advantages.

- Higher layer functionalities, such as transport and application layers, are processed at line rate, reducing latency and increasing throughput.
- Reduces traffic, thereby reducing network congestion, which is one of the factors degrading application performance.
- Saving energy by running servers [27].

First of all, it is important to determine what type of computation operations are most optimal to run in-network. According to the studies, the most feasible applications are in-network packet aggregation, in-network caching and applications alleviating control plane load [1]. Since data centres are early adopters of the SDN network, most of these applications are currently designed for data centre networks.

### 3.2.1.   In-Network Packet Aggregation

The group of applications with partition/aggregation patterns in the data centre network includes search, query processing, dataflow computing, graph processing, and stream processing, and deep learning frameworks. During the partitioning phase, job requests are divided into sub-tasks, which are executed in parallel on different worker servers and each worker server produces partial results. In order to obtain the final result, the partial results are collected and aggregated at the aggregation stage. During the aggregation phase, data (e.g partial results) must be transmitted between a large numbers of workers, which puts a heavy load on the network. For example, a trace of Facebook's data center shows that 46 percent of all traffic is generated during the aggregation phase. Furthermore, it leads to network bottleneck [28].

Data aggregation functionality is usually performed at the application layer. If it is done on the network path, traffic load can be reduced. Other reasons for in-network aggregations are that behind these functions are usually simple arithmetic logic operations, so placing them on a switch is simple, and since these algorithms are communicative associative functions, there is no need to pay attention to the packet sequence. DAIET [27] which is built in P4 is an in-network aggregation prototype solution for machine learning and graph analytics applications. However, the solution is generic enough and can be used in various partition/aggregation data centre application.

The next effective segment to use in-network aggregation is to combine small-sized and large numbers of packets. The idea has existed for a very long time, but there is no real implementation. This kind of packet aggregation/disaggregation has many important benefits. For example, aggregating multiple, and small-sized IoT packets into one transmission unit can reduce the additional overhead associated with each transfer. Wang et al. [29] introduce proof-of-concept designs and implementations on IoT packet aggregation and disaggregation purely in P4 pipelines of the switching ASIC.

### 3.2.2.   Applications alleviating control plane load

It is now technically possible to offload most of the tasks on a control plane to a data plane. The main benefit of this is that it can accelerate the control plane. However, some tasks are not optimal to run on data plane because they require a lot of resources. Therefore, research on what tasks should be offloaded on the data plane is one of the interesting topics in the future.

The case study on [30] suggested how to perform key control plane tasks such as failure detection, and notification, connectivity retrieval, and computation on a data plane, and implemented the proposed algorithms in BMv2 P4 software switch. It also discussed the advantages and disadvantages of implementing control plane tasks on a data plan. Another case study in this topic is the implementation of Time-synchronization Protocol (DPTP) on the Tofino programmable switch with P4 pipeline [30]. Global time-synchronization on the data-plane is very much necessary for supporting distributed applications. The key research questions around this are, first, to determine what types of control plane tasks can be optimally deployed on the data plane and how to overcome hardware constraints in deployment.

### 3.2.3.   In-network caching

Modern network services such as search engines, social networking e-commerce are used by billions of users and generate huge amounts of traffic on the network. To view a single web page, you may need to access hundreds or thousands of storage servers in the background [31]. One mechanism to deliver these services to users with high throughput and low latency is caching, which is a crucial way to improve the performance of a storage system. The idea is that to retrieve items on the storage system more quickly, high-access items (hot items) must be temporarily stored in the cache and the cache should be updated regularly based on hot items. The hot items can be changed abruptly, and most users like to access that hot items, which can lead to an imbalance in network traffic. For example, 60-70 percent of Facebook users access 10 percent of the total content [32]. Therefore, when building a caching system, these issues need to be considered.

Traditional networks use flash-based caches, disk-based caches, and server-based caches, and data plane programming provides new opportunity to create in-network caching. This means that it is possible to create a cache on a programmable network device. Because network devices naturally placed on the path between the client and server, creating a cache on the path can further reduce latency. The key-value store data structure is used to build the database in the cache because it is general and widely used by applications. It is used as a basic interface to the caching[33].

Netcache [34] is new key-value store architecture by leveraging flexibility, and the power of a modern programmable switch to handle queries on hot items of the storage server. It is built on top of rack (ToR) switch in the data centre network. Therefore, ToR switch plays important role and has 3 main modules: L2/L3 routing, on-path caching for key-value items, and query statistics. The Query statistic module identifies the hot items, and based on these statistics, the controller updates the cache. The core of Netcache is packet-processing pipeline which detect, index, store and serve key-value items. For example match-action table classify key on packet header and values are stored in register array, on-chip memory in programmable switch. One ToR switch can cache items on a storage server only connected to it, and cannot work with other ToR switches in a coherent way.

IncCache [35] is another in-network, key-value store system built in a programmable data plane. What distinguishes it from Netcache is that it is implemented in the core, aggregation and ToR switch of the data centre network, as well as end-host server, and maintains the cache coherence using a directory-based cache coherence protocol.

These works are good start for in-network caching and both reduce latency by a certain percentage. The Netcache architecture was created on a Tofino and commodity server-based switch with a P4 pipeline, while IncCache was developed on Cavium XPliant switch and the forwarding plane was defined in a proprietary language.

According to the discussion on those works, the following questions can be open in the future: Mostly network requests (read) are processed from the cache. So, can write/delete requests be processed from cache? Do you need compression to reduce the cache size? , and so on.

### 3.2.4. Consensus protocols (in-network)

Running some application-level protocols on the data plane is another interesting topic: for example, the implementation of consensus protocols for distributed networks in the data plane. Paxos is popular consensus protocol used in fault-tolerant networks and is commonly used in data center applications. Implementing this in the data plane will improve the performance of the protocol itself and the performance of applications based on this protocol service [36]. Data plane programmability allows for tight integration between the application and the network but, the developers should always consider how network-level optimization affects the top level.

## 3.3. Load Balancing Applications

The main purpose of the load balancer is to efficiently distribute the load over multiple pieces of network infrastructure in order to maximize throughput, minimize response time, and prevent overloading of single resource. The data centre network has redundant resources, so load balancers play an important role in the optimal use of these resources. Data centre networks perform load balancing in more than one way. The L3 load balancer(s) selects one of the many equal-cost paths that can route the packet, while the L4 load balancer(s) chooses the one of serving instances (servers) for the incoming service request [37].

Layer 3 load balancing mechanisms in the Data Centre network and Internet try to choose the congestion-free and optimal path from the multiple paths, so that bisection bandwidth can be used more efficiently. Layer 3 load balancing mechanisms are usually implemented on the data plane.

The most commonly implemented method is the Equal-Cost Multi-path Routing(ECMP) and the per-flow based load balancing mechanism, which randomly assigns one of the equal cost paths to each flow. Because the flow is distributed randomly, performance may be reduced if two elephant flows are allocated in the same path [38]. In addition, it is the congestion-oblivious mechanism that does not track the over-utilized path.

Conga [39] was improved ECMP, and it is a congestion-aware mechanism and maintains the congestion status of each path on the leaf/spine switch in the data centre network. However, due to the limited memory of the switch (leaf), it is not possible to scale this mechanism as the network grows. In addition, because Conga is implemented on custom hardware, it is costly to redesign (requires modification on chip architecture), which means that network operators cannot change the mechanism to suit their network.

With the advent of the programmable data plane, it became possible to develop a customizable load balancing mechanism on the programmable data plane. HULA [40], programmed in P4 is the first load balancing mechanism explicitly designed for the programmable switch architecture and it is scalable and congestion-aware. Conga centralizes the congestion track at one point (leaf switch), while HULA does it in a distributed manner. Each HULA switch maintains only the congestion state for the best next hop to reach the destination, not the congestion state for all paths to the destination. Therefore, each HULA switch makes a local decision when selecting a path, while the CONGA depends on the leaf switch. By tracking congestion in this distributed way, scalability is better than CONGA. In addition, it can automatically detect network failures. Therefore, this work inspires network operators to create a more optimal load balancing mechanism for their network in a programmable data plane.

The layer 4 load balancers could be hardware, cluster of servers and commodity server, and they are usually implemented on the commodity server in data center. It is also called software load balancers (SLB). Thanks to data plane programmability, they can also be developed on the switching ASIC. When designing a Layer 4 load balancer, the following two are important. First, the incoming connections to the servers must be very well-tuned to the bisection bandwidth of the physical network (uniform load distribution of the incoming connections across the network and servers). Second, providing per connection consistency (PCC): the ability to map packets belonging to the same connection to the same server, even if there are presence changes to the active servers and load balancers. But, meeting both these requirements at the same time has been an elusive goal [41].

It was not easy to ensure the PCC because the switching ASIC doesn't have enough memory to store a large number of connection states. However, with the continuous increase in memory size, it is possible to implement the L4 load balancer on the switch. The main advantage of implementing it on switch is that there is no additional software load balancer in between application traffic and application server. This allows balancing load at line rate. SilkRoad [35] was proposed as a load balancer on a programmable switching ASIC and implemented using 400 lines of P4 code. The performance measurement on SilkRoad show that it can balance 10 million connections at line speed.

SHELL [42] tried to implement a stateless load balancer on P4-NetFPGA programmable switch, and it is easier to deploy on a network device than a stateful solution. Moreover, SHELL is application-agnostic and load-awareness.

The above descriptions provide examples of implementing L3 and L4 load balancers on a programmable switch. Network traffic is constantly changing, so load balancing mechanisms need to be congestion-aware, dynamic, and with low latency. The results of empirical analysis of these implementations seem reasonable. In the future, the researchers can do an analytical study in terms of optimality and scalability on these in order to look for opportunities improving dynamic nature.

### 3.4. General factors for optimizing the data plane applications

Each application use case has its own optimization factors, which are described above. In this section, we described the general factors of optimization. What application will be on the data plane, and what kind of application optimization is needed generally depend on the type of network, such as local area network (LAN), wide area network (WAN), data centre network, industrial network, time-sensitive network, etc. For example, load balancing, traffic management, and congestion control applications play an important role in a data centre, but may not apply to other types of networks. Subsequent optimization factors may include network topology, traffic types, and so on. These are explained in a little more detail with the specific example as in the followings:

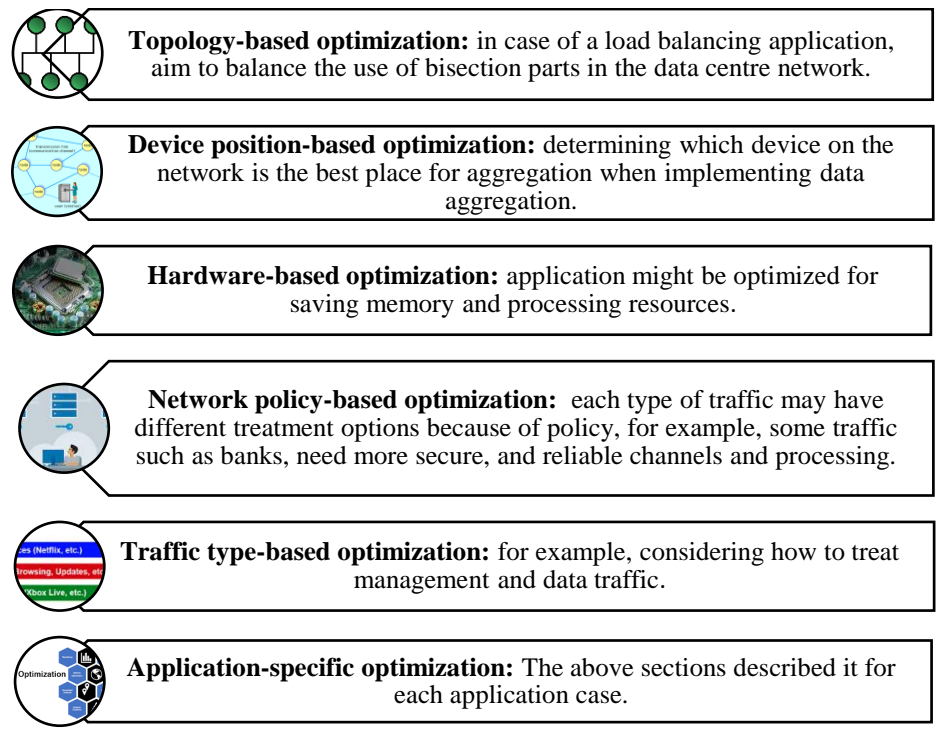

**Topology-based optimization:** in case of a load balancing application, aim to balance the use of bisection parts in the data centre network.

**Device position-based optimization:** determining which device on the network is the best place for aggregation when implementing data aggregation.

**Hardware-based optimization:** application might be optimized for saving memory and processing resources.

**Network policy-based optimization:** each type of traffic may have different treatment options because of policy, for example, some traffic such as banks, need more secure, and reliable channels and processing.

**Traffic type-based optimization:** for example, considering how to treat management and data traffic.

**Application-specific optimization:** The above sections described it for each application case.

Figure 4. Optimization factors

## 4. CONCLUSIONS AND FUTURE WORKS

With the advent of data plane programming, applications such as network monitoring, traffic aggregation, caching, and load balancing are being redesigned on the data plane. A lot of applications are developed on P4. Some solutions implemented on a BMv2 P4 prototype switch are not a guarantee that the solution will work effectively on real equipment in the production network, but it is a good start to design and promote innovation. Some are implemented on hardware switches such as FPGA and Tofino but have not been fully tested in the actual network. So there is a lot of work to be done to improve and optimize these solutions. However, it should be noted that the P4 application working group has developed some cases that can be used in the production network.

Many of these solutions are currently for data centers and promise to significantly improve the data center network. Similarly, there are many opportunities to develop new application cases in the data plane for other types of networks in the future. For example, the main requirement of an industrial network is reliability, low and predictable latency. Data plane programmability helps to reduce latency in the industrial network, and some prototype implementations such as in-network sensor monitoring, data caching for industrial automation, and in-network robotic control applications are made in BMv2 switch [43][44]. In addition, some solutions are emerging as the prototype for time-sensitive networks, such as in-network time synchronization. Furthermore, these kinds of studies can be well developed and optimized in the real network.

Also, it is possible to combine these applications and develop effective solutions. Therefore, the researchers can determine which combination of applications is the most optimal. For example, according to the INT specification, network troubleshooting, advanced congestion control, advanced routing, and network data plane verification can be made based on INT monitoring.

Other topics to consider about optimization are that developer needs to think how network optimization is related (or irrelevant) to transport-level optimization [45]. Current in-network works are mainly focused on optimizing the network layer. However, transport protocols will affect the performance of any in-network solution. In addition, programmable switches do not support floating point calculations used in more complex operations, such as artificial intelligence (AI) and machine learning (ML) algorithms. For example, AI-enabled analysis can be used to understand network problems caused by managing the complexity, scale, and dynamics of modern networks [46].

In the future, the development and optimization of a data plane program should take into account the general factors, and specific factors identified for each case used in our paper. This paper not only gave a better understanding of the data plane applications but also offered specific ideas about what can be done in the future.

## ACKNOWLEDGEMENTS

The research has been supported by the project "Application Domain Specific Highly Reliable IT Solutions" implemented with the support of the NRDI Fund of Hungary, financed under the Thematic Excellence Programme TKP2020-NKA-06 (National Challenges Sub programme) funding scheme.

## AUTHORS

**Tejfel Máté** received his B.Sc., M.Sc. and Ph.D. Degree in Computer Science, from ELTE, Budapest Hungary. He is currently working as an Associate Professor in the Department of Programming Languages and Compilers, ELTE. His research interest includes programming languages, correctness check, Software Defined Networks, and network optimization. For more information, visit his database at 0000-0001-8982-1398 orchid-id.

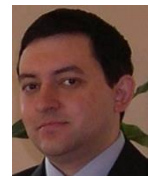

**Altangerel Gereltsetseg** received her B.Sc. & M.Sc. Degree in Information technology, from Mongolian University of Science and Technology (MUST). She is currently a Ph.D. student in the Department of Programming languages and compilers, ELTE under the supervision of Professor Tejfel Máté. Her research interest includes Software-defined Networks, deeply programmable network, and network optimization. For more information, visit her database at 0000-0002-1594-8158 orchid-id.

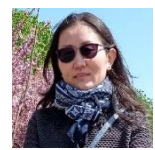